\documentclass[aip,amsmath,amssymb,
reprint]{revtex4-1}

\usepackage{graphicx}
\usepackage{dcolumn}
\usepackage{bm}

\usepackage[utf8]{inputenc}
\usepackage[T1]{fontenc}
\usepackage{mathptmx}
\usepackage{xcolor}

\begin{document}

\preprint{AIP/123-QED}

\title[]{Three-dimensional acoustic lensing with a bubbly diamond metamaterial\\}

\author{Maxime Lanoy}
 \email{maxime.lanoy@gmail.com.}

\affiliation{Laboratoire de Physique de l’ENS, Ecole Normale Supérieure, CNRS, Université de Paris, Sorbonne Université, 75005 Paris}
 
\author{Fabrice Lemoult}
\affiliation{Institut Langevin, ESPCI Paris, PSL University, CNRS, 75005 Paris}

\author{Geoffroy Lerosey}
\affiliation{Greenerwave, ESPCI Paris Incubator PC’up, 6 rue Jean Calvin, 75005 Paris, France}

\author{Arnaud Tourin}
\affiliation{Institut Langevin, ESPCI Paris, PSL University, CNRS, 75005 Paris}

\author{Valentin Leroy}
\affiliation{Laboratoire Matière et Systèmes Complexes, Université de Paris, CNRS, 75013 Paris}

\author{John H. Page}
\email{John.Page@UManitoba.ca}
\affiliation{Department of Physics and Astronomy, University of Manitoba, Winnipeg, Manitoba R3T 2N2, Canada}

\date{31 January 2021}

\begin{abstract}
A sound wave travelling in water is scattered by a periodic assembly of air bubbles. The local structure matters even in the low frequency regime. If the bubbles are arranged in a face-centered cubic (fcc) lattice, a total band gap opens near the Minnaert resonance frequency. If they are arranged in the diamond structure, which one obtains by simply adding a second bubble to the unit cell, one finds an additional branch with a negative slope (optical branch). For a single specific frequency, the medium behaves as if its refractive index (relative to water) is exactly $n=-1$. We show that a slab of this material can be used to design a three three-dimensional flat lens. We also report super-resolution focusing in the near field of the slab and illustrate its potential for imaging in three dimensions.

\end{abstract}

\maketitle


\section{\label{sec:level1}Introduction}\protect
All classical wave phenomena are influenced by a single quantity: the refractive index. In the late 1960s, the Russian researcher Veselago wondered how an electromagnetic wave would propagate if this index became negative\cite{veselago1967electrodynamics}. He predicted that, at the interface between a negative index material (NIM) and a regular medium, negative refraction would occur. Three decades later, the first experimental observation of a negative refractive index was performed for electromagnetic waves in the microwave range\cite{Shelby77}. In the previous year, 
Pendry\cite{pendry2000negative} had claimed that a slab made of a NIM of index $n=-1$ could be turned into a perfect lens, without being subject to the 
diffraction limit. A quest for NIMs started in the early 2000s.    

The main strategy to obtain these negative indices has involved 
building artificial materials, namely \emph{metamaterials}. In the low frequency regime, wave propagation can be fully described in terms of 
two macroscopically homogeneous effective parameters. For electromagnetic waves, the refractive index is negative when both the effective permittivity\cite{Pendry1998} and permeability\cite{Pendry1996,Pendry1999} are negative. For acoustic waves, the effective density\cite{Liu1734} and compressibility\cite{fang2006ultrasonic} should be simultaneously negative. Most designs rely on the use of subwavelength resonant scatterers, which may either be arranged in random or ordered configurations.  To achieve double negativity,  one can tune the effective parameters of the medium by exploiting the nature of these resonances. 
For instance, in acoustics, a monopolar resonance can lead to a negative compressibility while a dipolar one gives a negative density. Combining these two types of resonances  has allowed\cite{li2004double,brunet2015soft} a negative index, or strictly speaking a negative sound velocity, to be reached.

Interestingly, anomalous refraction, and notably a negative one, had actually been observed earlier in another kind of artificial medium, namely the \emph{photonic/phononic crystal}\cite{kosaka1998superprism,luo2002all,yang2004focusing,xu2013all,dubois2019acoustic}. Note that these media can be either two-\cite{kosaka1998superprism,luo2002all} or three-\cite{yang2004focusing,xu2013all,dubois2019acoustic} dimensional. 
The refraction in this case is actually due to the intrinsic periodic nature of these media, and can be seen as a generalization of the diffraction orders of periodic gratings. Indeed, due to the periodicity, the propagating eigensolutions inside these media are Bloch waves, which are the superposition of many plane waves. They all share the same orientation of the group velocity, which completely governs the refraction of beams. The implementation of this concept in a two-dimensional phononic crystal~\cite{sukhovich2008negative} has even reached the ultimate case of all angles being refracted similarly, thus resembling the case of an effective medium of index $n=-1$ . Focusing below the diffraction limit was also demonstrated in this case as a consequence of trapped slab modes\cite{sukhovich2009experimental,robillard2011resolution}. 

In between metamaterials with randomly arranged scatterers and phononic crystals 
resides a category of media that we could call \emph{crystalline metamaterials}\cite{kaina2015negative,lemoult_soda_2016,yves_left-handed_2019}. Indeed, close to the resonance frequency of the scatterers introduced inside a metamaterial, the effective wavelength can become as small as the average separation distance between scatterers. Then, 
the long wavelength approximation no longer applies and the propagation is sensitive to the spatial structure of the medium. This sensitivity to structure was a crucial aspect of the observation of negative refraction 
in two dimensional materials  in acoustics consisting of a honeycomb arrangement of soda cans\cite{kaina2015negative}. 

\begin{figure*}
\centering
\includegraphics[width=.7\textwidth]{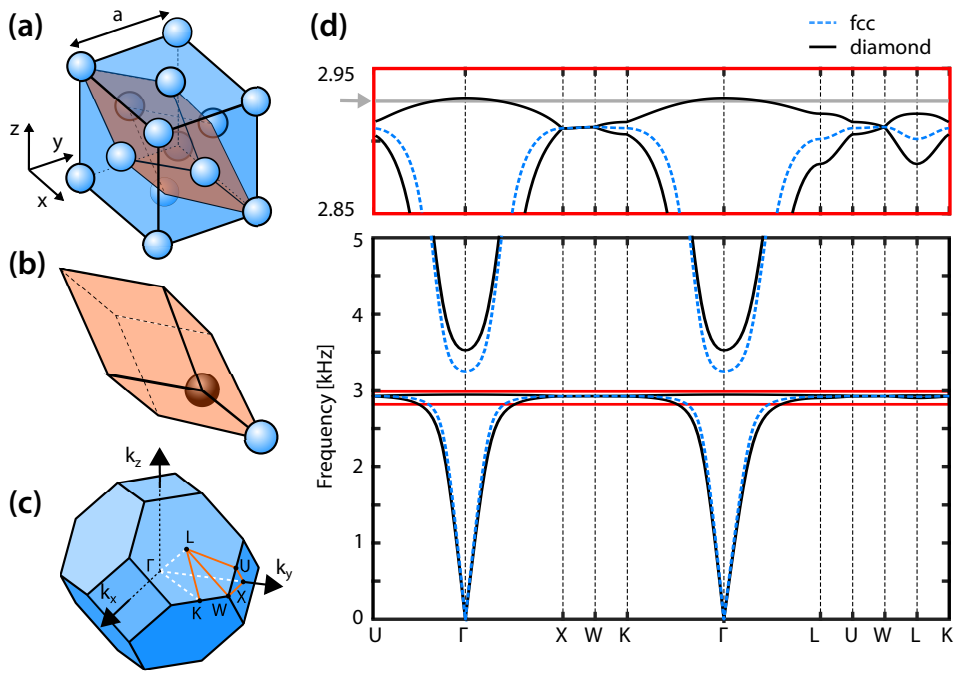}
\caption{(a) Sketch picturing the super-cell of a face-centered cubic (fcc) lattice populated with air bubbles. (b) Corresponding unit-cell in the fcc case (blue bubble) and in the diamond case (blue and brown bubbles). (c) First Brillouin zone for the fcc lattice. (d) Band structure for the fcc lattice (dashed blue line) and diamond structure (black line) computed with COMSOL Multiphysics by finding the eigenvalues for an infinite crystal (periodic boundary conditions). The upper plot corresponds to a zoom in the 2850 to 2950~Hz frequency range, with the solid grey horizontal line intersecting the optical branch of the diamond band structure at the frequency for which the $n=-1$ condition applies.}
\label{fig1}
\end{figure*}

In this article, we exploit a similar strategy to design a three-dimensional medium which exhibits all-angle negative refraction. The medium is made of underwater bubbles, which are effective ultrasonic low-frequency resonant scatterers. Interestingly, the appropriate three-dimensional crystalline structure turns out to be that of diamond, one of the natural crystals of the carbon atom (analogous to the honeycomb structure in two dimensions). We  verify numerically the negative refraction of a beam as well as the focusing  of a point source by a flat lens. We also demonstrate its capabilities in terms of subwavelength focusing. Note that the recent advances in 3D printed frames for the design of stable air bubble assemblies~\cite{choi2016large,cai2019bubble,combriat2020acoustic} make our proposal achievable in practice. Also, as only monopolar scatterers and the specific periodicity are relevant here, it makes the design adaptable to airborne acoustics (with spherical Helmholtz resonators for instance) or even to other kind of waves (\textit{e.g.} elastic waves).

\section{\label{sec:level1}Metamaterial design}\protect

As indicated in the previous paragraph, the design of our acoustic metamaterial is based on the acoustic resonance of air bubbles. Indeed, the significant contrast in density and in compressibility between the inclusion (air) and the host medium (water) results in a monopolar resonance that occurs at very low frequency: the so-called Minnaert resonance~\cite{minnaert1933xvi}. For instance, the resonance of an air bubble of radius $r=1$~mm is expected at $f_\text{m}=2890$~Hz\footnote{These calculations were done for air enriched with C$_6$F$_{14}$ vapor, due to the advantages for experiments of increased bubble stability in time, as well as reduced thermal losses. Consequently, the heat capacity ratio $\gamma=1.1$, which is less than the value for pure air and which results in a slight lowering of the bubble resonance frequency. }, which corresponds to a wavelength in water of 52~cm (520 times the bubble radius). In this regime, the pressure field at resonance is roughly homogeneous over the whole bubble surface and the acoustic response can be considered as isotropic. When a plane wave of unity amplitude impinges on such an air bubble, it gives rise to a scattered spherical wave whose amplitude $s(\omega)$ can therefore be modelled by a complex Lorentzian function:
\begin{equation}
s(\omega)=-\frac{r}{1-\omega_\text{m}^2/\omega^2+\text{i} k_0 r}.
\label{eq1}
\end{equation}
Here, $k_0$ stands for the wavenumber of the initial incident plane wave in bulk water and $\omega_\text{m}$ corresponds to the Minnaert frequency. Note that this expression does not account for viscous and thermal damping mechanisms, which will be discussed later on. 

When several bubbles coexist, multiple scattering effects must be taken into account~\cite{kafesaki2000air}. Although there is no analytical solution for this self-consistent problem, the calculation can be computed thanks to the formalism introduced by Lax~\cite{lax1952multiple}. Technical details are available in a previous publication dealing with multiple scattering of sound in a disordered bubbly medium~\cite{lanoy2015subwavelength}. At low frequency, the propagation in such a multiple scattering medium of bubbles is similar to the propagation in a homogeneous medium described by effective parameters which are essentially ruled by the scattering function $s(\omega)$ and the concentration of scatterers. %
In this article, instead of a random cloud, we propose to position the bubbles (radius $r=1$~mm) in a periodic fashion. We start off by 
arranging them on the lattice sites of a face-centered cubic (fcc) lattice [Fig.~\ref{fig1}(a)], with a lattice constant of $a=11.8$~cm (Note that this means the sample is highly diluted since the corresponding air volume fraction is $\phi=0.001\%$). This crystal is expected to behave roughly isotropically since its Brillouin zone is a fairly regular octahedron [see Fig.\ref{fig1}(c)], a characteristic that facilitated the first observation of a complete band gap in three-dimensional phononic crystals~\cite{yang2002ultrasound}. The eigenvalue problem is numerically solved in COMSOL considering the unit cell drawn in Fig.~\ref{fig1}(b) and applying periodic boundary conditions on its edges. Following this procedure along the main directions of the crystal, one obtains the band structure displayed in Fig.~\ref{fig1}(d) (blue solid line). One can notice two propagative bands (branches) separated by a complete band gap that opens near the Minnaert frequency. As for the random distribution of scatterers~\cite{lanoy2015subwavelength}, the propagation could be described in terms of an effective compressibility which is influenced by the resonance of the bubbles. The band gap corresponds to a negative compressibility while the flattening of the dispersion relation for the lower branch corresponds to a high compressibility. 

One interesting feature can be seen for the branch just below the bandgap, near the Minnaert frequency: it exhibits solutions for wavenumbers at the edges of the first Brillouin zone. This might seem surprising since, as pointed out earlier, the wavelength in water is 52~cm (\textit{i.e.}, nearly 5 times the lattice constant $a$). This indicates that, near this frequency, the sound wave travels inside the crystal with a much smaller ``effective'' wavelength, suggesting that both the local and long range spatial structure of the medium 
can influence the propagation even in the very low frequency regime.  Thus, judicious modification of the crystal structure could be expected to have a significant impact. We illustrate this effect by introducing a second bubble into the primitive unit cell so that it contains two identical scatterers with $x,y,z$ coordinates 0,0,0 and  {\footnotesize$\frac{1}{4}$},{\footnotesize$\frac{1}{4}$},{\footnotesize$\frac{1}{4}$} in units of the cubic cell parameter $a$.  We hereby obtain the \textit{diamond} structure, for which the corresponding band structure is displayed in Fig.~\ref{fig1}(d).  One consequence of adding the second bubble is that the nearest-neighbour spacing of the bubbles is reduced, with each pair of bubbles in the unit cell being separated by only $0.433a$, compared with a separation of $0.707a$ in the original fcc structure. While this increase in the bubble concentration is responsible for small changes in the band structure at low frequencies (\textit{e.g.}, the slope near zero frequency is slightly smaller, due to a rather modest increase in the effective compressibility of the diamond crystal), by far the most significant change in the band structure is the emergence of a new branch within the original fcc bandgap. Interestingly, this branch features a negative slope and can be regarded as a metamaterial equivalent of the optical branch observed in the phonon dispersion relations of diatomic crystals. 

\begin{figure}[h] 
    \centering
    \includegraphics[width=.45\textwidth]{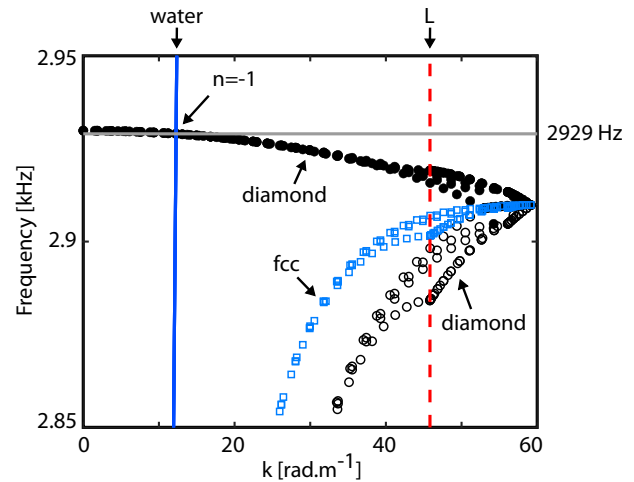}
    \caption{Dispersion relations obtained by plotting all branches from Fig.~\ref{fig1}(d) in a single graph for which the frequencies are displayed as a function of the magnitude of the wavevector for the different directions in the Brillouin zone.  Open symbols correspond to the lower frequency branches (blue for fcc, black for diamond), whereas solid black symbols depict the negative branch (only for diamond). The almost vertical blue line shows the dispersion in pure water and the red dashed line indicates the wavenumber magnitude at the L point.}
    \label{figS1}
\end{figure}

In addition to having a low-frequency negative branch, which is associated with the existence of two bubbles in the primitive unit cell, the metamaterial with the diamond structure has another notable feature, namely that propagation is expected to be isotropic, or nearly so. Figure~\ref{figS1} indicates how isotropic the behaviour actually is, and compares results for diamond with those of the fcc structure. Starting from the band structures of Fig.~\ref{fig1}(d), the normal mode frequencies for all branches and different directions are displayed in a single dispersion graph as a function of the wavevector magnitude. We find that all points for the different directions match up very well, especially for wavevectors in the lower two thirds of the Brillouin zone ($k \lesssim 40$ rad$\cdot$m$^{-1}$). Beyond the L point (vertical red dashed line), which corresponds to the shortest wavevector touching the Brillouin zone boundary, the spread of wavevector values at a given frequency increases  and the propagation becomes anisotropic (there is even a gap in certain directions). But returning back to lower wavevectors and focusing on the negative branch of the diamond structure (solid black circles in Fig.~\ref{figS1}), we note that the deviations from isotropy become less and less as the frequency increases, with the deviations appearing to be negligible when the dispersion curves cross the water line (solid blue line in Fig.~\ref{figS1}).  At the crossing frequency (2929 Hz), the variation in the magnitude of the wavevectors is less than 0.3$\%$, so that the equifrequency surface is essentially  spherical and the propagation really is for all practical purposes isotropic.  

\begin{figure}[bt]
    \centering
    \includegraphics[width=.35\textwidth]{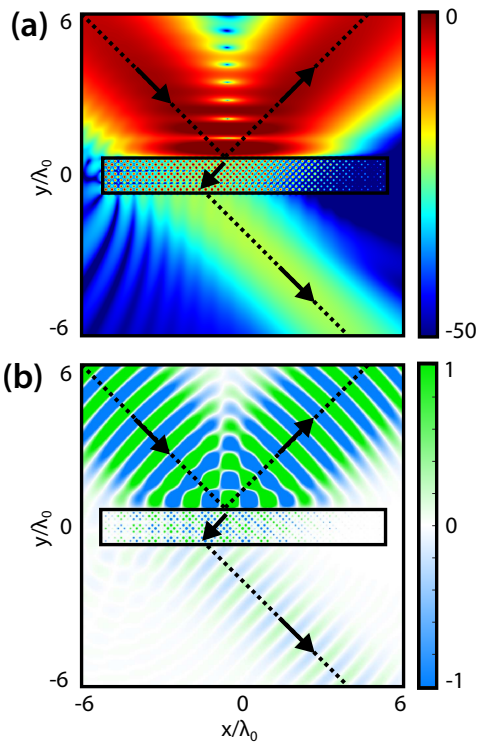}
    \caption{Refraction of an obliquely incident Gaussian beam on a slab of bubbly diamond. The arrows indicate the propagation directions. (a) Normalized energy in dB. (b) Real part of the pressure field, showing the wavefronts. Here, $\lambda_0=51.2$~cm.}
    \label{fig2}
\end{figure}

\section{\label{sec:level1}Results and Discussion}
\subsection{\label{sec:level2}Negative refraction}
For the optical branch, as we increase the frequency, the radius of the isofrequency surface decreases, meaning that this branch is associated with a negative refraction effect: an obliquely incident wavevector is refracted through the crystal interface with a negative angle. Here, we illustrate the negative refraction by shining a Gaussian beam towards the bubbly diamond metamaterial at an incident angle of 45$^{\circ}$ with respect to the normal at the interface. Using the method explained in reference 29
, we compute the full solution to the multiple scattering problem of the acoustic wave field inside the metamaterial when this beam is  incident on a slab of thickness $1.5\lambda_0$ and width 
$11\lambda_0$ containing 81,000 bubbles.  (Here, $\lambda_0$ denotes the wavelength in water.)  In Fig.~\ref{fig2}(a), we display the corresponding acoustic energy map (in dB) at the frequency 2929 Hz. The incident beam is mostly reflected at the first interface but the remaining energy which enters the crystal is negatively refracted at the two slab interfaces. It finally exits the slab with a spatial shift towards negative $x$ values. The real part of the pressure field, shown in Fig.~\ref{fig2}(b), provides additional information. First, looking at the wave-fronts' orientation inside the crystal, we find that the refraction angle is exactly opposite to the incident angle, suggesting a phase velocity being the exact opposite of the water one. This is further confirmed by the fact that the wavelengths inside and outside the crystal are equal at this frequency. Nevertheless, we have to stress that the $n=-1$ condition is not accompanied by good impedance matching, as shown by the significant reflection at the first interface. 

\begin{figure}[bt]
    \centering
    \includegraphics[width=.35\textwidth]{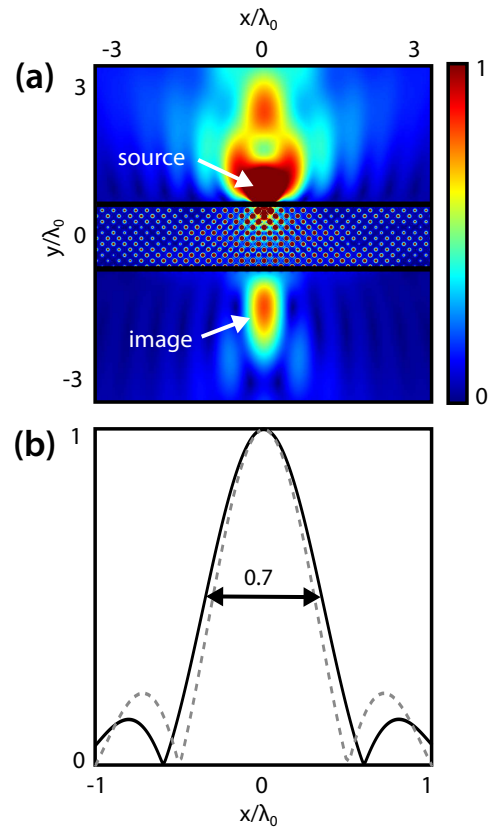}
    \caption{(a) Map of the pressure field magnitude obtained when the bubbly diamond slab is excited by a point source facing its upper edge. At a frequency (2929 Hz) such that $n=-1$, the wave-vectors are collected in phase at a conjugate location on the opposite slide of the slab thus forming an image of the source. (b) Transverse profile of the focal spot (solid line) and comparison with an optimal diffraction-limited focus (dashed line). Here, $\lambda_0=51.2$~cm.}
    \label{fig3}
\end{figure}

\subsection{\label{sec:level2}Focusing of a point source}
Since the original predictions of Veselago~\cite{veselago1967electrodynamics} and Pendry~\cite{pendry2000negative}, $n=-1$ materials have been attracting much attention for the possibilities they offer in terms of imaging. The seminal configuration consists of a point source placed  in front of a slab with a $n=-1$ refractive index. The expectation is that all wavevectors emitted by the source spontaneously converge back to a single location on the other side of the slab. The only requirement for the formation of an image on the far side of the slab is that the source should not be located further than the slab thickness away from the first slab's interface. Here, the point source is located just outside the slab, in front of a bubble, with the distance from the source to the bubble being approximately $\lambda_0/17$.   We numerically investigate this configuration and observe the focusing effect through the slab [see Fig.~\ref{fig3}(a)] by again solving the multiple scattering problem of wave transport inside the slab, this time resulting from a point source excitation. To estimate the lateral resolution, the  transverse profile at the focus is displayed as a solid line in Fig.~\ref{fig3}(b). In this case, the distance from the peak to the first minimum, $\Delta/2$, is $0.6\lambda_0$, and the corresponding Full Width at Half Maximum (FWHM) is $0.7\lambda_0$.  This is close to the optimal value set by the Rayleigh criterion, which considers diffraction-related limitations. 
It basically states that wavevectors of magnitude higher than $\omega/c$ (with $c$ being the acoustic wave velocity) cannot be collected simply because these waves are evanescent and cannot propagate in the uniform medium outside the slab, since they decrease exponentially away from the source. 
\begin{figure}
    \centering
    \includegraphics[width=.35\textwidth]{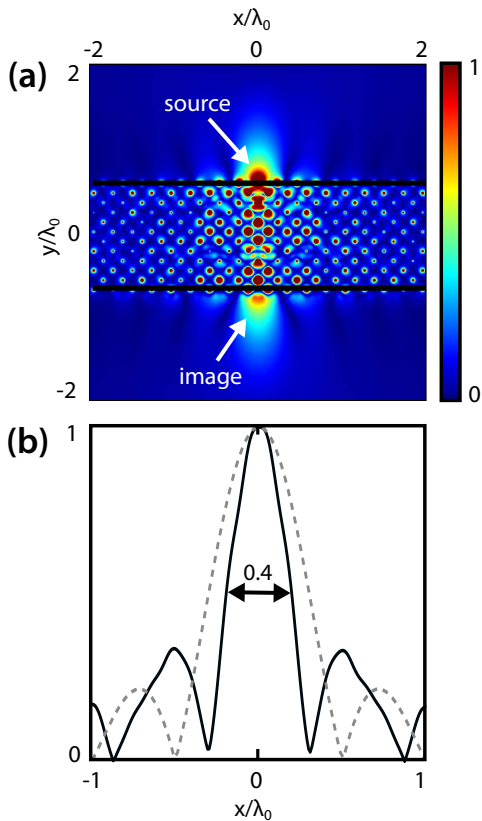}
    \caption{(a) Map of the pressure field magnitude obtained when the slab is excited by a point source facing its upper edge. At a frequency (2917 Hz) such that $n=-3.2$, the wave-vectors are collected in phase at a conjugate location on the opposite side of the slab, thus forming an image of the source. Since the image is located in the near field, evanescent wavevectors are involved in the image formation. (b) Transverse profile of the focal spot (solid line) and comparison with an optimal diffraction-limited focus (dashed line). Here, $\lambda_0=51.4$~cm.}
    \label{fig4}
\end{figure}

However, if the image of the point source is located in the near field of the lens, the evanescent waves at these larger wavevectors are not eliminated by the need to propagate out of the slab. As a consequence, the information they carry can be collected at the focal point. In order to shift the focus closer to the slab, one can simply increase the refractive index of the metalens which, it turns out, can be done by slightly lowering the frequency. For instance, at 2917~Hz the refractive index is roughly $n=-3.2$ which sets the focus at a distance $\lambda_0/10$ from the bottom interface [see Fig.~\ref{fig4}(a)]. Again, we display the transverse profile of the focal spot in Fig.~\ref{fig4}(b); the FWHM is $0.4\lambda_0$ and $\Delta/2=0.3\lambda_0$, which beats the lower limit predicted by the Rayleigh criterion.\\

\begin{figure*}
    \centering
    \includegraphics[width=.90\textwidth]{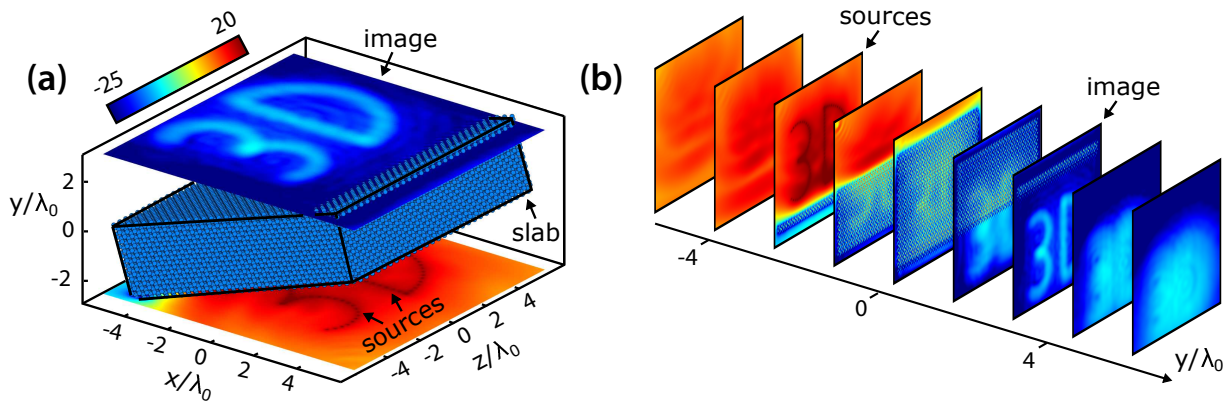}
    \caption{(a) Imaging in three dimensions of an inclined extended planar object consisting of the letters 3D. Note that the colorbar scale is expressed in dB. (b) Slices of the acoustic energy (in dB) in equally spaced planes perpendicular to the $y$-axis in (a), illustrating the depth of field of the image.}
    \label{fig5}
\end{figure*}

\subsection{\label{sec:level2}Imaging in three dimensions}
The focusing effect for a point source is just a step towards demonstrating that the device is capable of imaging an actual extended object. Our bubbly metalens operates in three dimensions. To highlight this specific attribute, we now use the lens to image an extended source consisting of 92 incoherent individual point sources arranged to form the acronym "3D" [see Fig.~\ref{fig5}(a)]. The object is actually two-dimensional and contained 
in the ($x$,$z$) plane. In order to demonstrate the ability of the lens to image an object that extends over a three-dimensional region of space near the lens, we simply tilt the slab around the $z$-axis (by $18^{\circ}$).  
(This, of course, is equivalent to keeping the lens fixed in position while rotating the object by $-18^{\circ}$.)  Tilting the slab (or alternatively tilting the planar object) amounts to breaking rotational invariance with respect to the $y$-axis, thereby involving all three dimensions in the imaging process. For the image shown in Fig.~\ref{fig5}(a), the bubble concentration in the slab was reduced by a factor of 4, thereby extending the lateral width of the lens to enable a larger object to be imaged without excessively impacting the computation time.  The frequency is 2910~Hz, which corresponds to $n=-1$ for this diluted concentration. After propagation through the slab, an image that respects the shape, size and orientation of the original object is created [Fig.~\ref{fig5}(a)].  To illustrate the depth of field of the image, two-dimensional cross sections of the acoustic energy in equally spaced planes perpendicular to the $y$-axis are displayed in Fig.~\ref{fig5}(b).  

\subsection{\label{sec:level2}The damping issue}
So far, damping terms resulting from thermal and viscous mechanisms have not been considered. As a consequence, it is not obvious if similar results would be observed in a realistic bubble slab. More specifically, with the chosen set of parameters, these additional terms actually undermine the lensing effect because the field decays faster. As the characteristic decay length becomes comparable to the slab width, fewer wavevectors are collected at the location of the focus simply because obliquely incident wavevectors cover a longer distance within the slab before reaching the second interface. As a result they undergo more scattering events and are hence more attenuated than their normally incident counterparts. However, there are concrete solutions to this limitation, allowing it to be overcome. 
In a previous contribution reporting a negative branch in a random set of pair-wise correlated bubbles, we showed that increasing the bubble concentration was a means of reducing attenuation in the frequency range of the negative branch~\cite{lanoy2017acoustic}. Although we could not examine higher concentrations here due to practical computation limitations, we expect similar trends considering the strong analogy between the diamond bi-periodic structure and the disordered pair-correlated system. 

Another possible way of reducing the overall dissipation is to consider using bigger bubbles. Indeed both the thermal and viscous damping terms weaken as the bubble radius is increased~\cite{devin1959survey}. In practice, air bubbles in water that are much larger than a millimeter in diameter cannot be considered, as their shapes deviate from spherical. However, the trapping technique recently developed by Harazi \textit{et al.}~\cite{harazi2019acoustics} allows much bigger ranges of size to be explored, since the shape is then set by a solid frame. Finally, one should keep in mind that the Lorentzian formalism of Eq.~(\ref{eq1}) is quite universal among scattering problems. As a result, our results are transferable to other kinds of metamaterials. The only real requirement 
concerns the scattering function which should remain isotropic. Three-dimensional spherical Helmholtz resonators could be envisaged as these objects tend to exhibit less loss. 

\section{\label{sec:level1}Conclusions}\protect
In this article, we demonstrate numerically that a bubble assembly organized in the diamond structure constitutes an effective three-dimensional acoustic flat metalens. The negative refractive index in the optical branch of this crystalline metamaterial's dispersion relation is a consequence of the bi-periodic structure, and is highlighted by demonstrating the anomalous refraction of an obliquely incident Gaussian beam. The diamond structure consists of a fcc Bravais lattice with a basis of two scatterers per unit cell. Just like the fcc lattice, it is highly isotropic, as shown in Fig.~\ref{figS1}: 
the refractive index is essentially the same in all directions throughout a large part of the Brillouin zone, with  deviations from isotropy only becoming clearly noticeable when the equifrequency surfaces approach 
the zone boundary, which they do first at the L points (where the  Brillouin zone edges are closest to the zone center).  Isotropic propagation is a key feature to avoid aberration effects, making the choice of the diamond structure for this three-dimensional metalens an ideal one compared with other crystalline structures that might be envisaged.  
We demonstrate the lensing performances by presenting evidence of the focusing of a single point source. The position of the focus can be tuned by sweeping the frequencies. As the focus moves into the near-field of the slab, the focus gets sharper and eventually beats the Rayleigh criterion. This indicates that the evanescent field then contributes to the image formation so that the imaging resolution is not limited to half the operating wavelength. Finally, imaging of an extended object is demonstrated to illustrate the three-dimensional possibilities.\\

Beyond the specific properties of this diamond structure, there is an interesting physics point to make here. This study should be regarded as a crystalline counterpart of our previous work on the consequences of introducing a spatial pair correlation within a disordered assembly of monopolar resonators~\cite{lanoy2017acoustic}. In this previous paper, we considered pairs of monopolar scatterers, separated by a fixed distance, that were otherwise randomly distributed in the medium; the resulting pair correlation was found to induce a statistically resilient dipolar resonance. At the dipolar resonance frequency, the effective density and the effective refractive index both become negative (the effective bulk modulus being negative is a consequence of the monopolar resonance). In the diamond structure metamaterial, the physics is intrinsically analogous. The bi-periodicity due to the presence of two scatterers in the primitive unit cell can be regarded as a way of introducing a dipolar mode. However, since the observations could very well be interpreted with strictly crystallographic arguments, no mention was made previously in this article to such a dipolar mode or to the effective density.  This brings forward a fascinating feature of the metamaterial regime. The intensive parameters of the material can be dictated by the properties at the scale of the basic building block of the structure (\emph{e.g.}, unit cell in the crystalline case or nearest-neighbor separation for a random distribution of bubble pairs) and essentially rely on two aspects. The first is the nature of the inclusion (monopole, dipole...) while the second is its spatial patterning (random, pair-correlated, fcc, diamond...). This rather general statement is illustrated here by the mechanism leading to a negative branch in the dispersion relation of two disparate structures built from bubble pairs:  on the one hand we can consider that the bubble pair is a single complex scatterer with two resonances that may be positioned randomly in the medium~\cite{lanoy2017acoustic}, while on the other hand we have a spatially relevant crystalline combination of two identical monopolar scatterers (the description presented in this article). Thus, under certain circumstances, the macroscopic properties of metamaterials that might normally be considered to be very different (e.g., disordered dispersions of complex scatterers versus perfectly ordered bi-periodic crystals) can end up being two facets of a single phenomenon.

\begin{acknowledgments}
Some of this research has been supported by LABEX WIFI (Laboratory of Excellence within the French Program "Investments for the Future") under references ANR-10-LABX-24 and ANR-10-IDEX-0001-02 PSL, and by Agence Nationale de la Recherche under reference ANR-16-CE31-0015.  JHP would like to acknowledge support from the Natural Sciences and Engineering Research Council of Canada’s Discovery Grant Program (RGPIN-2016-06042).
\end{acknowledgments}

\section*{\label{sec:level1}Data availability}
The data that supports the findings of this study are available within the article

\section*{\label{sec:level1}References}

\begin{thebibliography}{33}%
\makeatletter
\providecommand \@ifxundefined [1]{%
 \@ifx{#1\undefined}
}%
\providecommand \@ifnum [1]{%
 \ifnum #1\expandafter \@firstoftwo
 \else \expandafter \@secondoftwo
 \fi
}%
\providecommand \@ifx [1]{%
 \ifx #1\expandafter \@firstoftwo
 \else \expandafter \@secondoftwo
 \fi
}%
\providecommand \natexlab [1]{#1}%
\providecommand \enquote  [1]{``#1''}%
\providecommand \bibnamefont  [1]{#1}%
\providecommand \bibfnamefont [1]{#1}%
\providecommand \citenamefont [1]{#1}%
\providecommand \href@noop [0]{\@secondoftwo}%
\providecommand \href [0]{\begingroup \@sanitize@url \@href}%
\providecommand \@href[1]{\@@startlink{#1}\@@href}%
\providecommand \@@href[1]{\endgroup#1\@@endlink}%
\providecommand \@sanitize@url [0]{\catcode `\\12\catcode `\$12\catcode
  `\&12\catcode `\#12\catcode `\^12\catcode `\_12\catcode `\%12\relax}%
\providecommand \@@startlink[1]{}%
\providecommand \@@endlink[0]{}%
\providecommand \url  [0]{\begingroup\@sanitize@url \@url }%
\providecommand \@url [1]{\endgroup\@href {#1}{\urlprefix }}%
\providecommand \urlprefix  [0]{URL }%
\providecommand \Eprint [0]{\href }%
\providecommand \doibase [0]{http://dx.doi.org/}%
\providecommand \selectlanguage [0]{\@gobble}%
\providecommand \bibinfo  [0]{\@secondoftwo}%
\providecommand \bibfield  [0]{\@secondoftwo}%
\providecommand \translation [1]{[#1]}%
\providecommand \BibitemOpen [0]{}%
\providecommand \bibitemStop [0]{}%
\providecommand \bibitemNoStop [0]{.\EOS\space}%
\providecommand \EOS [0]{\spacefactor3000\relax}%
\providecommand \BibitemShut  [1]{\csname bibitem#1\endcsname}%
\let\auto@bib@innerbib\@empty
\bibitem [{\citenamefont {Veselago}(1967)}]{veselago1967electrodynamics}%
  \BibitemOpen
  \bibfield  {author} {\bibinfo {author} {\bibfnamefont {V.~G.}\ \bibnamefont
  {Veselago}},\ }\bibfield  {title} {\enquote {\bibinfo {title}
  {Electrodynamics of substances with simultaneously negative and},}\
  }\href@noop {} {\bibfield  {journal} {\bibinfo  {journal} {Usp. Fiz. Nauk.}\
  }\textbf {\bibinfo {volume} {92}},\ \bibinfo {pages} {517} (\bibinfo {year}
  {1967})}\BibitemShut {NoStop}%
\bibitem [{\citenamefont {Shelby}, \citenamefont {Smith},\ and\ \citenamefont
  {Schultz}(2001)}]{Shelby77}%
  \BibitemOpen
  \bibfield  {author} {\bibinfo {author} {\bibfnamefont {R.~A.}\ \bibnamefont
  {Shelby}}, \bibinfo {author} {\bibfnamefont {D.~R.}\ \bibnamefont {Smith}}, \
  and\ \bibinfo {author} {\bibfnamefont {S.}~\bibnamefont {Schultz}},\
  }\bibfield  {title} {\enquote {\bibinfo {title} {Experimental verification of
  a negative index of refraction},}\ }\href {\doibase 10.1126/science.1058847}
  {\bibfield  {journal} {\bibinfo  {journal} {Science}\ }\textbf {\bibinfo
  {volume} {292}},\ \bibinfo {pages} {77--79} (\bibinfo {year}
  {2001})}\BibitemShut {NoStop}%
\bibitem [{\citenamefont {Pendry}(2000)}]{pendry2000negative}%
  \BibitemOpen
  \bibfield  {author} {\bibinfo {author} {\bibfnamefont {J.~B.}\ \bibnamefont
  {Pendry}},\ }\bibfield  {title} {\enquote {\bibinfo {title} {Negative
  refraction makes a perfect lens},}\ }\href@noop {} {\bibfield  {journal}
  {\bibinfo  {journal} {Phys. Rev. Lett.}\ }\textbf {\bibinfo {volume} {85}},\
  \bibinfo {pages} {3966} (\bibinfo {year} {2000})}\BibitemShut {NoStop}%
\bibitem [{\citenamefont {Pendry}\ \emph {et~al.}(1998)\citenamefont {Pendry},
  \citenamefont {Holden}, \citenamefont {Robbins},\ and\ \citenamefont
  {Stewart}}]{Pendry1998}%
  \BibitemOpen
  \bibfield  {author} {\bibinfo {author} {\bibfnamefont {J.~B.}\ \bibnamefont
  {Pendry}}, \bibinfo {author} {\bibfnamefont {A.}~\bibnamefont {Holden}},
  \bibinfo {author} {\bibfnamefont {D.}~\bibnamefont {Robbins}}, \ and\
  \bibinfo {author} {\bibfnamefont {W.}~\bibnamefont {Stewart}},\ }\bibfield
  {title} {\enquote {\bibinfo {title} {Low frequency plasmons in thin-wire
  structures},}\ }\href@noop {} {\bibfield  {journal} {\bibinfo  {journal} {J.
  Condens. Matter Phys.}\ }\textbf {\bibinfo {volume} {10}},\ \bibinfo {pages}
  {4785} (\bibinfo {year} {1998})}\BibitemShut {NoStop}%
\bibitem [{\citenamefont {Pendry}\ \emph {et~al.}(1996)\citenamefont {Pendry},
  \citenamefont {Holden}, \citenamefont {Stewart},\ and\ \citenamefont
  {Youngs}}]{Pendry1996}%
  \BibitemOpen
  \bibfield  {author} {\bibinfo {author} {\bibfnamefont {J.~B.}\ \bibnamefont
  {Pendry}}, \bibinfo {author} {\bibfnamefont {A.}~\bibnamefont {Holden}},
  \bibinfo {author} {\bibfnamefont {W.}~\bibnamefont {Stewart}}, \ and\
  \bibinfo {author} {\bibfnamefont {I.}~\bibnamefont {Youngs}},\ }\bibfield
  {title} {\enquote {\bibinfo {title} {Extremely low frequency plasmons in
  metallic mesostructures},}\ }\href@noop {} {\bibfield  {journal} {\bibinfo
  {journal} {Phys. Rev. Lett.}\ }\textbf {\bibinfo {volume} {76}},\ \bibinfo
  {pages} {4773} (\bibinfo {year} {1996})}\BibitemShut {NoStop}%
\bibitem [{\citenamefont {Pendry}\ \emph {et~al.}(1999)\citenamefont {Pendry},
  \citenamefont {Holden}, \citenamefont {Robbins},\ and\ \citenamefont
  {Stewart}}]{Pendry1999}%
  \BibitemOpen
  \bibfield  {author} {\bibinfo {author} {\bibfnamefont {J.~B.}\ \bibnamefont
  {Pendry}}, \bibinfo {author} {\bibfnamefont {A.~J.}\ \bibnamefont {Holden}},
  \bibinfo {author} {\bibfnamefont {D.~J.}\ \bibnamefont {Robbins}}, \ and\
  \bibinfo {author} {\bibfnamefont {W.}~\bibnamefont {Stewart}},\ }\bibfield
  {title} {\enquote {\bibinfo {title} {Magnetism from conductors and enhanced
  nonlinear phenomena},}\ }\href@noop {} {\bibfield  {journal} {\bibinfo
  {journal} {IEEE Trans. Microw. Theory Tech.}\ }\textbf {\bibinfo {volume}
  {47}},\ \bibinfo {pages} {2075--2084} (\bibinfo {year} {1999})}\BibitemShut
  {NoStop}%
\bibitem [{\citenamefont {Liu}\ \emph {et~al.}(2000)\citenamefont {Liu},
  \citenamefont {Zhang}, \citenamefont {Mao}, \citenamefont {Zhu},
  \citenamefont {Yang}, \citenamefont {Chan},\ and\ \citenamefont
  {Sheng}}]{Liu1734}%
  \BibitemOpen
  \bibfield  {author} {\bibinfo {author} {\bibfnamefont {Z.}~\bibnamefont
  {Liu}}, \bibinfo {author} {\bibfnamefont {X.}~\bibnamefont {Zhang}}, \bibinfo
  {author} {\bibfnamefont {Y.}~\bibnamefont {Mao}}, \bibinfo {author}
  {\bibfnamefont {Y.~Y.}\ \bibnamefont {Zhu}}, \bibinfo {author} {\bibfnamefont
  {Z.}~\bibnamefont {Yang}}, \bibinfo {author} {\bibfnamefont {C.~T.}\
  \bibnamefont {Chan}}, \ and\ \bibinfo {author} {\bibfnamefont
  {P.}~\bibnamefont {Sheng}},\ }\bibfield  {title} {\enquote {\bibinfo {title}
  {Locally resonant sonic materials},}\ }\href {\doibase
  10.1126/science.289.5485.1734} {\bibfield  {journal} {\bibinfo  {journal}
  {Science}\ }\textbf {\bibinfo {volume} {289}},\ \bibinfo {pages} {1734--1736}
  (\bibinfo {year} {2000})}\BibitemShut {NoStop}%
\bibitem [{\citenamefont {Fang}\ \emph {et~al.}(2006)\citenamefont {Fang},
  \citenamefont {Xi}, \citenamefont {Xu}, \citenamefont {Ambati}, \citenamefont
  {Srituravanich}, \citenamefont {Sun},\ and\ \citenamefont
  {Zhang}}]{fang2006ultrasonic}%
  \BibitemOpen
  \bibfield  {author} {\bibinfo {author} {\bibfnamefont {N.}~\bibnamefont
  {Fang}}, \bibinfo {author} {\bibfnamefont {D.}~\bibnamefont {Xi}}, \bibinfo
  {author} {\bibfnamefont {J.}~\bibnamefont {Xu}}, \bibinfo {author}
  {\bibfnamefont {M.}~\bibnamefont {Ambati}}, \bibinfo {author} {\bibfnamefont
  {W.}~\bibnamefont {Srituravanich}}, \bibinfo {author} {\bibfnamefont
  {C.}~\bibnamefont {Sun}}, \ and\ \bibinfo {author} {\bibfnamefont
  {X.}~\bibnamefont {Zhang}},\ }\bibfield  {title} {\enquote {\bibinfo {title}
  {Ultrasonic metamaterials with negative modulus},}\ }\href@noop {} {\bibfield
   {journal} {\bibinfo  {journal} {Nat. Mater.}\ }\textbf {\bibinfo {volume}
  {5}},\ \bibinfo {pages} {452--456} (\bibinfo {year} {2006})}\BibitemShut
  {NoStop}%
\bibitem [{\citenamefont {Li}\ and\ \citenamefont {Chan}(2004)}]{li2004double}%
  \BibitemOpen
  \bibfield  {author} {\bibinfo {author} {\bibfnamefont {J.}~\bibnamefont
  {Li}}\ and\ \bibinfo {author} {\bibfnamefont {C.~T.}\ \bibnamefont {Chan}},\
  }\bibfield  {title} {\enquote {\bibinfo {title} {Double-negative acoustic
  metamaterial},}\ }\href@noop {} {\bibfield  {journal} {\bibinfo  {journal}
  {Phys. Rev. E}\ }\textbf {\bibinfo {volume} {70}},\ \bibinfo {pages} {055602}
  (\bibinfo {year} {2004})}\BibitemShut {NoStop}%
\bibitem [{\citenamefont {Brunet}\ \emph {et~al.}(2015)\citenamefont {Brunet},
  \citenamefont {Merlin}, \citenamefont {Mascaro}, \citenamefont {Zimny},
  \citenamefont {Leng}, \citenamefont {Poncelet}, \citenamefont
  {Arist{\'e}gui},\ and\ \citenamefont {Mondain-Monval}}]{brunet2015soft}%
  \BibitemOpen
  \bibfield  {author} {\bibinfo {author} {\bibfnamefont {T.}~\bibnamefont
  {Brunet}}, \bibinfo {author} {\bibfnamefont {A.}~\bibnamefont {Merlin}},
  \bibinfo {author} {\bibfnamefont {B.}~\bibnamefont {Mascaro}}, \bibinfo
  {author} {\bibfnamefont {K.}~\bibnamefont {Zimny}}, \bibinfo {author}
  {\bibfnamefont {J.}~\bibnamefont {Leng}}, \bibinfo {author} {\bibfnamefont
  {O.}~\bibnamefont {Poncelet}}, \bibinfo {author} {\bibfnamefont
  {C.}~\bibnamefont {Arist{\'e}gui}}, \ and\ \bibinfo {author} {\bibfnamefont
  {O.}~\bibnamefont {Mondain-Monval}},\ }\bibfield  {title} {\enquote {\bibinfo
  {title} {Soft 3d acoustic metamaterial with negative index},}\ }\href@noop {}
  {\bibfield  {journal} {\bibinfo  {journal} {Nat. Mater.}\ }\textbf {\bibinfo
  {volume} {14}},\ \bibinfo {pages} {384--388} (\bibinfo {year}
  {2015})}\BibitemShut {NoStop}%
\bibitem [{\citenamefont {Kosaka}\ \emph {et~al.}(1998)\citenamefont {Kosaka},
  \citenamefont {Kawashima}, \citenamefont {Tomita}, \citenamefont {Notomi},
  \citenamefont {Tamamura}, \citenamefont {Sato},\ and\ \citenamefont
  {Kawakami}}]{kosaka1998superprism}%
  \BibitemOpen
  \bibfield  {author} {\bibinfo {author} {\bibfnamefont {H.}~\bibnamefont
  {Kosaka}}, \bibinfo {author} {\bibfnamefont {T.}~\bibnamefont {Kawashima}},
  \bibinfo {author} {\bibfnamefont {A.}~\bibnamefont {Tomita}}, \bibinfo
  {author} {\bibfnamefont {M.}~\bibnamefont {Notomi}}, \bibinfo {author}
  {\bibfnamefont {T.}~\bibnamefont {Tamamura}}, \bibinfo {author}
  {\bibfnamefont {T.}~\bibnamefont {Sato}}, \ and\ \bibinfo {author}
  {\bibfnamefont {S.}~\bibnamefont {Kawakami}},\ }\bibfield  {title} {\enquote
  {\bibinfo {title} {Superprism phenomena in photonic crystals},}\ }\href@noop
  {} {\bibfield  {journal} {\bibinfo  {journal} {Phys. Rev. B}\ }\textbf
  {\bibinfo {volume} {58}},\ \bibinfo {pages} {R10096} (\bibinfo {year}
  {1998})}\BibitemShut {NoStop}%
\bibitem [{\citenamefont {Luo}\ \emph {et~al.}(2002)\citenamefont {Luo},
  \citenamefont {Johnson}, \citenamefont {Joannopoulos},\ and\ \citenamefont
  {Pendry}}]{luo2002all}%
  \BibitemOpen
  \bibfield  {author} {\bibinfo {author} {\bibfnamefont {C.}~\bibnamefont
  {Luo}}, \bibinfo {author} {\bibfnamefont {S.~G.}\ \bibnamefont {Johnson}},
  \bibinfo {author} {\bibfnamefont {J.}~\bibnamefont {Joannopoulos}}, \ and\
  \bibinfo {author} {\bibfnamefont {J.}~\bibnamefont {Pendry}},\ }\bibfield
  {title} {\enquote {\bibinfo {title} {All-angle negative refraction without
  negative effective index},}\ }\href@noop {} {\bibfield  {journal} {\bibinfo
  {journal} {Phys. Rev. B}\ }\textbf {\bibinfo {volume} {65}},\ \bibinfo
  {pages} {201104} (\bibinfo {year} {2002})}\BibitemShut {NoStop}%
\bibitem [{\citenamefont {Yang}\ \emph {et~al.}(2004)\citenamefont {Yang},
  \citenamefont {Page}, \citenamefont {Liu}, \citenamefont {Cowan},
  \citenamefont {Chan},\ and\ \citenamefont {Sheng}}]{yang2004focusing}%
  \BibitemOpen
  \bibfield  {author} {\bibinfo {author} {\bibfnamefont {S.}~\bibnamefont
  {Yang}}, \bibinfo {author} {\bibfnamefont {J.~H.}\ \bibnamefont {Page}},
  \bibinfo {author} {\bibfnamefont {Z.}~\bibnamefont {Liu}}, \bibinfo {author}
  {\bibfnamefont {M.~L.}\ \bibnamefont {Cowan}}, \bibinfo {author}
  {\bibfnamefont {C.~T.}\ \bibnamefont {Chan}}, \ and\ \bibinfo {author}
  {\bibfnamefont {P.}~\bibnamefont {Sheng}},\ }\bibfield  {title} {\enquote
  {\bibinfo {title} {Focusing of sound in a 3d phononic crystal},}\ }\href@noop
  {} {\bibfield  {journal} {\bibinfo  {journal} {Phys. Rev. Lett.}\ }\textbf
  {\bibinfo {volume} {93}},\ \bibinfo {pages} {024301} (\bibinfo {year}
  {2004})}\BibitemShut {NoStop}%
\bibitem [{\citenamefont {Xu}\ \emph {et~al.}(2013)\citenamefont {Xu},
  \citenamefont {Agrawal}, \citenamefont {Abashin}, \citenamefont {Chau},\ and\
  \citenamefont {Lezec}}]{xu2013all}%
  \BibitemOpen
  \bibfield  {author} {\bibinfo {author} {\bibfnamefont {T.}~\bibnamefont
  {Xu}}, \bibinfo {author} {\bibfnamefont {A.}~\bibnamefont {Agrawal}},
  \bibinfo {author} {\bibfnamefont {M.}~\bibnamefont {Abashin}}, \bibinfo
  {author} {\bibfnamefont {K.~J.}\ \bibnamefont {Chau}}, \ and\ \bibinfo
  {author} {\bibfnamefont {H.~J.}\ \bibnamefont {Lezec}},\ }\bibfield  {title}
  {\enquote {\bibinfo {title} {All-angle negative refraction and active flat
  lensing of ultraviolet light},}\ }\href@noop {} {\bibfield  {journal}
  {\bibinfo  {journal} {Nature}\ }\textbf {\bibinfo {volume} {497}},\ \bibinfo
  {pages} {470--474} (\bibinfo {year} {2013})}\BibitemShut {NoStop}%
\bibitem [{\citenamefont {Dubois}\ \emph {et~al.}(2019)\citenamefont {Dubois},
  \citenamefont {Perchoux}, \citenamefont {Vanel}, \citenamefont {Tronche},
  \citenamefont {Achaoui}, \citenamefont {Dupont}, \citenamefont {Bertling},
  \citenamefont {Raki{\'c}}, \citenamefont {Antonakakis}, \citenamefont
  {Enoch}, \citenamefont {Abdeddaim}, \citenamefont {Craster},\ and\
  \citenamefont {Guenneau}}]{dubois2019acoustic}%
  \BibitemOpen
  \bibfield  {author} {\bibinfo {author} {\bibfnamefont {M.}~\bibnamefont
  {Dubois}}, \bibinfo {author} {\bibfnamefont {J.}~\bibnamefont {Perchoux}},
  \bibinfo {author} {\bibfnamefont {A.}~\bibnamefont {Vanel}}, \bibinfo
  {author} {\bibfnamefont {C.}~\bibnamefont {Tronche}}, \bibinfo {author}
  {\bibfnamefont {Y.}~\bibnamefont {Achaoui}}, \bibinfo {author} {\bibfnamefont
  {G.}~\bibnamefont {Dupont}}, \bibinfo {author} {\bibfnamefont
  {K.}~\bibnamefont {Bertling}}, \bibinfo {author} {\bibfnamefont
  {A.}~\bibnamefont {Raki{\'c}}}, \bibinfo {author} {\bibfnamefont
  {T.}~\bibnamefont {Antonakakis}}, \bibinfo {author} {\bibfnamefont
  {S.}~\bibnamefont {Enoch}}, \bibinfo {author} {\bibfnamefont
  {R.}~\bibnamefont {Abdeddaim}}, \bibinfo {author} {\bibfnamefont {R.~V.}\
  \bibnamefont {Craster}}, \ and\ \bibinfo {author} {\bibfnamefont
  {S.}~\bibnamefont {Guenneau}},\ }\bibfield  {title} {\enquote {\bibinfo
  {title} {Acoustic flat lensing using an indefinite medium},}\ }\href@noop {}
  {\bibfield  {journal} {\bibinfo  {journal} {Phys. Rev. B}\ }\textbf {\bibinfo
  {volume} {99}},\ \bibinfo {pages} {100301} (\bibinfo {year}
  {2019})}\BibitemShut {NoStop}%
\bibitem [{\citenamefont {Sukhovich}, \citenamefont {Jing},\ and\ \citenamefont
  {Page}(2008)}]{sukhovich2008negative}%
  \BibitemOpen
  \bibfield  {author} {\bibinfo {author} {\bibfnamefont {A.}~\bibnamefont
  {Sukhovich}}, \bibinfo {author} {\bibfnamefont {L.}~\bibnamefont {Jing}}, \
  and\ \bibinfo {author} {\bibfnamefont {J.~H.}\ \bibnamefont {Page}},\
  }\bibfield  {title} {\enquote {\bibinfo {title} {Negative refraction and
  focusing of ultrasound in two-dimensional phononic crystals},}\ }\href@noop
  {} {\bibfield  {journal} {\bibinfo  {journal} {Phys. Rev. E}\ }\textbf
  {\bibinfo {volume} {77}},\ \bibinfo {pages} {014301} (\bibinfo {year}
  {2008})}\BibitemShut {NoStop}%
\bibitem [{\citenamefont {Sukhovich}\ \emph {et~al.}(2009)\citenamefont
  {Sukhovich}, \citenamefont {Merheb}, \citenamefont {Muralidharan},
  \citenamefont {Vasseur}, \citenamefont {Pennec}, \citenamefont {Deymier},\
  and\ \citenamefont {Page}}]{sukhovich2009experimental}%
  \BibitemOpen
  \bibfield  {author} {\bibinfo {author} {\bibfnamefont {A.}~\bibnamefont
  {Sukhovich}}, \bibinfo {author} {\bibfnamefont {B.}~\bibnamefont {Merheb}},
  \bibinfo {author} {\bibfnamefont {K.}~\bibnamefont {Muralidharan}}, \bibinfo
  {author} {\bibfnamefont {J.~O.}~\bibnamefont {Vasseur}}, \bibinfo {author}
  {\bibfnamefont {Y.}~\bibnamefont {Pennec}}, \bibinfo {author} {\bibfnamefont
  {P.~A.}\ \bibnamefont {Deymier}}, \ and\ \bibinfo {author} {\bibfnamefont
  {J.~H.}\ \bibnamefont {Page}},\ }\bibfield  {title} {\enquote {\bibinfo
  {title} {Experimental and theoretical evidence for subwavelength imaging in
  phononic crystals},}\ }\href@noop {} {\bibfield  {journal} {\bibinfo
  {journal} {Phys. Rev. Lett.}\ }\textbf {\bibinfo {volume} {102}},\ \bibinfo
  {pages} {154301} (\bibinfo {year} {2009})}\BibitemShut {NoStop}%
\bibitem [{\citenamefont {Robillard}\ \emph {et~al.}(2011)\citenamefont
  {Robillard}, \citenamefont {Bucay}, \citenamefont {Deymier}, \citenamefont
  {Shelke}, \citenamefont {Muralidharan}, \citenamefont {Merheb}, \citenamefont
  {Vasseur}, \citenamefont {Sukhovich},\ and\ \citenamefont
  {Page}}]{robillard2011resolution}%
  \BibitemOpen
  \bibfield  {author} {\bibinfo {author} {\bibfnamefont {J.-F.}~\ \bibnamefont
  {Robillard}}, \bibinfo {author} {\bibfnamefont {J.}~\bibnamefont {Bucay}},
  \bibinfo {author} {\bibfnamefont {P.~A.}~\bibnamefont {Deymier}}, \bibinfo
  {author} {\bibfnamefont {A.}~\bibnamefont {Shelke}}, \bibinfo {author}
  {\bibfnamefont {K.}~\bibnamefont {Muralidharan}}, \bibinfo {author}
  {\bibfnamefont {B.}~\bibnamefont {Merheb}}, \bibinfo {author} {\bibfnamefont
  {J.~O.}~\bibnamefont {Vasseur}}, \bibinfo {author} {\bibfnamefont
  {A.}~\bibnamefont {Sukhovich}}, \ and\ \bibinfo {author} {\bibfnamefont
  {J.~H.}~\ \bibnamefont {Page}},\ }\bibfield  {title} {\enquote {\bibinfo
  {title} {Resolution limit of a phononic crystal superlens},}\ }\href@noop {}
  {\bibfield  {journal} {\bibinfo  {journal} {Phys. Rev. B}\ }\textbf {\bibinfo
  {volume} {83}},\ \bibinfo {pages} {224301} (\bibinfo {year}
  {2011})}\BibitemShut {NoStop}%
\bibitem [{\citenamefont {Kaina}\ \emph {et~al.}(2015)\citenamefont {Kaina},
  \citenamefont {Lemoult}, \citenamefont {Fink},\ and\ \citenamefont
  {Lerosey}}]{kaina2015negative}%
  \BibitemOpen
  \bibfield  {author} {\bibinfo {author} {\bibfnamefont {N.}~\bibnamefont
  {Kaina}}, \bibinfo {author} {\bibfnamefont {F.}~\bibnamefont {Lemoult}},
  \bibinfo {author} {\bibfnamefont {M.}~\bibnamefont {Fink}}, \ and\ \bibinfo
  {author} {\bibfnamefont {G.}~\bibnamefont {Lerosey}},\ }\bibfield  {title}
  {\enquote {\bibinfo {title} {Negative refractive index and acoustic superlens
  from multiple scattering in single negative metamaterials},}\ }\href@noop {}
  {\bibfield  {journal} {\bibinfo  {journal} {Nature}\ }\textbf {\bibinfo
  {volume} {525}},\ \bibinfo {pages} {77--81} (\bibinfo {year}
  {2015})}\BibitemShut {NoStop}%
\bibitem [{\citenamefont {Lemoult}\ \emph {et~al.}(2016)\citenamefont
  {Lemoult}, \citenamefont {Kaina}, \citenamefont {Fink},\ and\ \citenamefont
  {Lerosey}}]{lemoult_soda_2016}%
  \BibitemOpen
  \bibfield  {author} {\bibinfo {author} {\bibfnamefont {F.}~\bibnamefont
  {Lemoult}}, \bibinfo {author} {\bibfnamefont {N.}~\bibnamefont {Kaina}},
  \bibinfo {author} {\bibfnamefont {M.}~\bibnamefont {Fink}}, \ and\ \bibinfo
  {author} {\bibfnamefont {G.}~\bibnamefont {Lerosey}},\ }\bibfield  {title}
  {\enquote {\bibinfo {title} {Soda cans metamaterial: A subwavelength-scaled
  phononic crystal},}\ }\href@noop {} {\bibfield  {journal} {\bibinfo
  {journal} {Crystals}\ }\textbf {\bibinfo {volume} {6}},\ \bibinfo {pages}
  {82} (\bibinfo {year} {2016})}\BibitemShut {NoStop}%
\bibitem [{\citenamefont {Yves}\ \emph {et~al.}(2019)\citenamefont {Yves},
  \citenamefont {Berthelot}, \citenamefont {Fink}, \citenamefont {Lerosey},\
  and\ \citenamefont {Lemoult}}]{yves_left-handed_2019}%
  \BibitemOpen
  \bibfield  {author} {\bibinfo {author} {\bibfnamefont {S.}~\bibnamefont
  {Yves}}, \bibinfo {author} {\bibfnamefont {T.}~\bibnamefont {Berthelot}},
  \bibinfo {author} {\bibfnamefont {M.}~\bibnamefont {Fink}}, \bibinfo {author}
  {\bibfnamefont {G.}~\bibnamefont {Lerosey}}, \ and\ \bibinfo {author}
  {\bibfnamefont {F.}~\bibnamefont {Lemoult}},\ }\bibfield  {title} {\enquote
  {\bibinfo {title} {Left-handed band in an electromagnetic metamaterial
  induced by sub-wavelength multiple scattering},}\ }\href@noop {} {\bibfield
  {journal} {\bibinfo  {journal} {Appl. Phys. Lett.}\ }\textbf {\bibinfo
  {volume} {114}},\ \bibinfo {pages} {111101} (\bibinfo {year}
  {2019})}\BibitemShut {NoStop}%
\bibitem [{\citenamefont {Choi}\ \emph {et~al.}(2016)\citenamefont {Choi},
  \citenamefont {David}, \citenamefont {Gao}, \citenamefont {Chang},
  \citenamefont {Allen}, \citenamefont {Wang},\ and\ \citenamefont
  {Chang}}]{choi2016large}%
  \BibitemOpen
  \bibfield  {author} {\bibinfo {author} {\bibfnamefont {C.-H.}\ \bibnamefont
  {Choi}}, \bibinfo {author} {\bibfnamefont {M.}~\bibnamefont {David}},
  \bibinfo {author} {\bibfnamefont {Z.}~\bibnamefont {Gao}}, \bibinfo {author}
  {\bibfnamefont {A.}~\bibnamefont {Chang}}, \bibinfo {author} {\bibfnamefont
  {M.}~\bibnamefont {Allen}}, \bibinfo {author} {\bibfnamefont
  {H.}~\bibnamefont {Wang}}, \ and\ \bibinfo {author} {\bibfnamefont {C.-h.}\
  \bibnamefont {Chang}},\ }\bibfield  {title} {\enquote {\bibinfo {title}
  {Large-scale generation of patterned bubble arrays on printed bi-functional
  boiling surfaces},}\ }\href@noop {} {\bibfield  {journal} {\bibinfo
  {journal} {Sci. Rep.}\ }\textbf {\bibinfo {volume} {6}},\ \bibinfo {pages}
  {23760} (\bibinfo {year} {2016})}\BibitemShut {NoStop}%
\bibitem [{\citenamefont {Cai}\ \emph {et~al.}(2019)\citenamefont {Cai},
  \citenamefont {Zhao}, \citenamefont {Huang}, \citenamefont {Li},
  \citenamefont {Su}, \citenamefont {Zhang}, \citenamefont {Zhao},
  \citenamefont {Hu}, \citenamefont {Wang},\ and\ \citenamefont
  {Song}}]{cai2019bubble}%
  \BibitemOpen
  \bibfield  {author} {\bibinfo {author} {\bibfnamefont {Z.}~\bibnamefont
  {Cai}}, \bibinfo {author} {\bibfnamefont {S.}~\bibnamefont {Zhao}}, \bibinfo
  {author} {\bibfnamefont {Z.}~\bibnamefont {Huang}}, \bibinfo {author}
  {\bibfnamefont {Z.}~\bibnamefont {Li}}, \bibinfo {author} {\bibfnamefont
  {M.}~\bibnamefont {Su}}, \bibinfo {author} {\bibfnamefont {Z.}~\bibnamefont
  {Zhang}}, \bibinfo {author} {\bibfnamefont {Z.}~\bibnamefont {Zhao}},
  \bibinfo {author} {\bibfnamefont {X.}~\bibnamefont {Hu}}, \bibinfo {author}
  {\bibfnamefont {Y.-S.}\ \bibnamefont {Wang}}, \ and\ \bibinfo {author}
  {\bibfnamefont {Y.}~\bibnamefont {Song}},\ }\bibfield  {title} {\enquote
  {\bibinfo {title} {Bubble architectures for locally resonant acoustic
  metamaterials},}\ }\href@noop {} {\bibfield  {journal} {\bibinfo  {journal}
  {Adv. Funct. Mater.}\ }\textbf {\bibinfo {volume} {29}},\ \bibinfo {pages}
  {1906984} (\bibinfo {year} {2019})}\BibitemShut {NoStop}%
\bibitem [{\citenamefont {Combriat}\ \emph {et~al.}(2020)\citenamefont
  {Combriat}, \citenamefont {Rouby-Poizat}, \citenamefont {Doinikov},
  \citenamefont {Stephan},\ and\ \citenamefont
  {Marmottant}}]{combriat2020acoustic}%
  \BibitemOpen
  \bibfield  {author} {\bibinfo {author} {\bibfnamefont {T.}~\bibnamefont
  {Combriat}}, \bibinfo {author} {\bibfnamefont {P.}~\bibnamefont
  {Rouby-Poizat}}, \bibinfo {author} {\bibfnamefont {A.~A.}\ \bibnamefont
  {Doinikov}}, \bibinfo {author} {\bibfnamefont {O.}~\bibnamefont {Stephan}}, \
  and\ \bibinfo {author} {\bibfnamefont {P.}~\bibnamefont {Marmottant}},\
  }\bibfield  {title} {\enquote {\bibinfo {title} {Acoustic interaction between
  3d-fabricated cubic bubbles},}\ }\href@noop {} {\bibfield  {journal}
  {\bibinfo  {journal} {Soft Matter}\ }\textbf {\bibinfo {volume} {16}},\
  \bibinfo {pages} {2829--2835} (\bibinfo {year} {2020})}\BibitemShut {NoStop}%
\bibitem [{\citenamefont {Minnaert}(1933)}]{minnaert1933xvi}%
  \BibitemOpen
  \bibfield  {author} {\bibinfo {author} {\bibfnamefont {M.}~\bibnamefont
  {Minnaert}},\ }\bibfield  {title} {\enquote {\bibinfo {title} {On musical
  air-bubbles and the sounds of running water},}\ }\href@noop {} {\bibfield
  {journal} {\bibinfo  {journal} {Philos. Mag}\ }\textbf {\bibinfo {volume}
  {16}},\ \bibinfo {pages} {235--248} (\bibinfo {year} {1933})}\BibitemShut
  {NoStop}%
\bibitem [{Note1()}]{Note1}%
  \BibitemOpen
  \bibinfo {note} {These calculations were done for air enriched with
  C$_6$F$_{14}$ vapor, due to the advantages for experiments of increased
  bubble stability in time, as well as reduced thermal losses. Consequently,
  the heat capacity ratio $\gamma =1.1$, which is less than the value for pure
  air and which results in a slight lowering of the bubble resonance
  frequency.}\BibitemShut {Stop}%
\bibitem [{\citenamefont {Kafesaki}, \citenamefont {Penciu},\ and\
  \citenamefont {Economou}(2000)}]{kafesaki2000air}%
  \BibitemOpen
  \bibfield  {author} {\bibinfo {author} {\bibfnamefont {M.}~\bibnamefont
  {Kafesaki}}, \bibinfo {author} {\bibfnamefont {R.}~\bibnamefont {Penciu}}, \
  and\ \bibinfo {author} {\bibfnamefont {E.}~\bibnamefont {Economou}},\
  }\bibfield  {title} {\enquote {\bibinfo {title} {Air bubbles in water: a
  strongly multiple scattering medium for acoustic waves},}\ }\href@noop {}
  {\bibfield  {journal} {\bibinfo  {journal} {Phys. Rev. Lett.}\ }\textbf
  {\bibinfo {volume} {84}},\ \bibinfo {pages} {6050} (\bibinfo {year}
  {2000})}\BibitemShut {NoStop}%
\bibitem [{\citenamefont {Lax}(1952)}]{lax1952multiple}%
  \BibitemOpen
  \bibfield  {author} {\bibinfo {author} {\bibfnamefont {M.}~\bibnamefont
  {Lax}},\ }\bibfield  {title} {\enquote {\bibinfo {title} {Multiple scattering
  of waves. ii. the effective field in dense systems},}\ }\href@noop {}
  {\bibfield  {journal} {\bibinfo  {journal} {Phys. Rev.}\ }\textbf {\bibinfo
  {volume} {85}},\ \bibinfo {pages} {621} (\bibinfo {year} {1952})}\BibitemShut
  {NoStop}%
\bibitem [{\citenamefont {Lanoy}\ \emph {et~al.}(2015)\citenamefont {Lanoy},
  \citenamefont {Pierrat}, \citenamefont {Lemoult}, \citenamefont {Fink},
  \citenamefont {Leroy},\ and\ \citenamefont
  {Tourin}}]{lanoy2015subwavelength}%
  \BibitemOpen
  \bibfield  {author} {\bibinfo {author} {\bibfnamefont {M.}~\bibnamefont
  {Lanoy}}, \bibinfo {author} {\bibfnamefont {R.}~\bibnamefont {Pierrat}},
  \bibinfo {author} {\bibfnamefont {F.}~\bibnamefont {Lemoult}}, \bibinfo
  {author} {\bibfnamefont {M.}~\bibnamefont {Fink}}, \bibinfo {author}
  {\bibfnamefont {V.}~\bibnamefont {Leroy}}, \ and\ \bibinfo {author}
  {\bibfnamefont {A.}~\bibnamefont {Tourin}},\ }\bibfield  {title} {\enquote
  {\bibinfo {title} {Subwavelength focusing in bubbly media using broadband
  time reversal},}\ }\href@noop {} {\bibfield  {journal} {\bibinfo  {journal}
  {Phys. Rev. B}\ }\textbf {\bibinfo {volume} {91}},\ \bibinfo {pages} {224202}
  (\bibinfo {year} {2015})}\BibitemShut {NoStop}%
\bibitem [{\citenamefont {Yang}\ \emph {et~al.}(2002)\citenamefont {Yang},
  \citenamefont {Page}, \citenamefont {Liu}, \citenamefont {Cowan},
  \citenamefont {Chan},\ and\ \citenamefont {Sheng}}]{yang2002ultrasound}%
  \BibitemOpen
  \bibfield  {author} {\bibinfo {author} {\bibfnamefont {S.}~\bibnamefont
  {Yang}}, \bibinfo {author} {\bibfnamefont {J.~H.}\ \bibnamefont {Page}},
  \bibinfo {author} {\bibfnamefont {Z.}~\bibnamefont {Liu}}, \bibinfo {author}
  {\bibfnamefont {M.~L.}\ \bibnamefont {Cowan}}, \bibinfo {author}
  {\bibfnamefont {C.~T.}\ \bibnamefont {Chan}}, \ and\ \bibinfo {author}
  {\bibfnamefont {P.}~\bibnamefont {Sheng}},\ }\bibfield  {title} {\enquote
  {\bibinfo {title} {Ultrasound tunneling through 3d phononic crystals},}\
  }\href@noop {} {\bibfield  {journal} {\bibinfo  {journal} {Phys. Rev. Lett.}\
  }\textbf {\bibinfo {volume} {88}},\ \bibinfo {pages} {104301} (\bibinfo
  {year} {2002})}\BibitemShut {NoStop}%
\bibitem [{\citenamefont {Lanoy}\ \emph {et~al.}(2017)\citenamefont {Lanoy},
  \citenamefont {Page}, \citenamefont {Lerosey}, \citenamefont {Lemoult},
  \citenamefont {Tourin},\ and\ \citenamefont {Leroy}}]{lanoy2017acoustic}%
  \BibitemOpen
  \bibfield  {author} {\bibinfo {author} {\bibfnamefont {M.}~\bibnamefont
  {Lanoy}}, \bibinfo {author} {\bibfnamefont {J.~H.}\ \bibnamefont {Page}},
  \bibinfo {author} {\bibfnamefont {G.}~\bibnamefont {Lerosey}}, \bibinfo
  {author} {\bibfnamefont {F.}~\bibnamefont {Lemoult}}, \bibinfo {author}
  {\bibfnamefont {A.}~\bibnamefont {Tourin}}, \ and\ \bibinfo {author}
  {\bibfnamefont {V.}~\bibnamefont {Leroy}},\ }\bibfield  {title} {\enquote
  {\bibinfo {title} {Acoustic double negativity induced by position
  correlations within a disordered set of monopolar resonators},}\ }\href@noop
  {} {\bibfield  {journal} {\bibinfo  {journal} {Phys. Rev. B}\ }\textbf
  {\bibinfo {volume} {96}},\ \bibinfo {pages} {220201} (\bibinfo {year}
  {2017})}\BibitemShut {NoStop}%
\bibitem [{\citenamefont {Devin~Jr}(1959)}]{devin1959survey}%
  \BibitemOpen
  \bibfield  {author} {\bibinfo {author} {\bibfnamefont {C.}~\bibnamefont
  {Devin~Jr}},\ }\bibfield  {title} {\enquote {\bibinfo {title} {Survey of
  thermal, radiation, and viscous damping of pulsating air bubbles in water},}\
  }\href@noop {} {\bibfield  {journal} {\bibinfo  {journal} {J. Acoust. Soc.
  Am.}\ }\textbf {\bibinfo {volume} {31}},\ \bibinfo {pages} {1654--1667}
  (\bibinfo {year} {1959})}\BibitemShut {NoStop}%
\bibitem [{\citenamefont {Harazi}\ \emph {et~al.}(2019)\citenamefont {Harazi},
  \citenamefont {Rupin}, \citenamefont {Stephan}, \citenamefont {Bossy},\ and\
  \citenamefont {Marmottant}}]{harazi2019acoustics}%
  \BibitemOpen
  \bibfield  {author} {\bibinfo {author} {\bibfnamefont {M.}~\bibnamefont
  {Harazi}}, \bibinfo {author} {\bibfnamefont {M.}~\bibnamefont {Rupin}},
  \bibinfo {author} {\bibfnamefont {O.}~\bibnamefont {Stephan}}, \bibinfo
  {author} {\bibfnamefont {E.}~\bibnamefont {Bossy}}, \ and\ \bibinfo {author}
  {\bibfnamefont {P.}~\bibnamefont {Marmottant}},\ }\bibfield  {title}
  {\enquote {\bibinfo {title} {Acoustics of cubic bubbles: six coupled
  oscillators},}\ }\href@noop {} {\bibfield  {journal} {\bibinfo  {journal}
  {Phys. Rev. Lett.}\ }\textbf {\bibinfo {volume} {123}},\ \bibinfo {pages}
  {254501} (\bibinfo {year} {2019})}\BibitemShut {NoStop}%
\end{thebibliography}

\providecommand{\noopsort}[1]{}\providecommand{\singleletter}[1]{#1}%

\end{document}